\documentstyle[twoside,fleqn,espcrc2]{article}
\catcode`\@=11 
\newcommand{\lsim}{\mathrel{\mathpalette\@versim<}}
\newcommand{\gsim}{\mathrel{\mathpalette\@versim>}}
\def\@versim#1#2{\vcenter{\offinterlineskip
 \ialign{$\m@th#1\hfil##\hfil$\crcr#2\crcr\sim\crcr } }}

\catcode`\@=12 

\input epsf
\ifx\epsffile\undefined
\def\epsfxsize#1{}
\def\epsffile#1{}
\else
\fi

\newcommand{\tg}{{\tilde g}}
\def\decay#1{\hbox{$\vrule height #1pt depth
0pt$}\kern-1pt\lower2.5pt\hbox{$\rightarrow$}}
\def\et{\not{\hbox{\kern-4.3pt $E_T$}}}

\title{Supersymmetry at LHC and NLC
\thanks{Work supported by the U.S.  National
Science Foundation, grant NSF-PHY-9404057.}
}

\author{Jonathan A.  Bagger
\address{Department of Physics and Astronomy,
 Johns Hopkins University \\
 3400 N.~Charles Street,
 Baltimore, MD 21218}}
 
\begin{document}

\begin{abstract}
\noindent
This talk discusses the prospects for supersymmetry studies at the LHC
and NLC.  The results are based on those of the Supersymmetry Working
Group at the 1996 Snowmass Workshop.
\end{abstract}

\maketitle

\section{INTRODUCTION}

For years, we supersymmeters have had to bear a heavy burden.  We
have been forced to defend the preposterous proposition that nature is
supersymmetric -- even though not one superpartner has been discovered.
Precision measurements \cite{EWWG} have eliminated much of the competition,
but they have done nothing to silence the popular press \cite{NYT}.
(For an alternate view, see \cite{Lane,Snowmass:TC}.)

Fortunately, this situation will soon change.  The upcoming generation
of supercolliders will settle the question once and for all.  The CERN
Large Hadron Collider (LHC), a 14 TeV $pp$ collider, is scheduled for
completion in 2005.  The Next Linear Collider (NLC), an expandable,
 0.5 to 1.5 TeV $e^+e^-$ linear collider, is proposed to begin operation
shortly thereafter.  These two machines will probe the entire parameter
space of weak-scale supersymmetry.  

The bottom line is that in less than a decade, our wait will be over.
We will finally know whether weak-scale supersymmetry is realized
in nature.  At SUSY07, we will either celebrate one of the greatest
triumphs in the history of science -- or join with the sociologists
in a searching discussion of how we could have gone so wrong
\cite{social}.

If supersymmetry is right, SUSY07 will mark the dawn of a new
era.  Theorists and experimentalists alike
will be beginning to explore a rich set of new physics.  Of
course, much time and hard work will be required to demonstrate that
the new physics is, in fact, supersymmetry.  In this talk I will
report on some preliminary steps taken in this direction by the
Supersymmetry Working Group at the 1996 Snowmass Workshop \cite{Snowmass}.  
This group examined the supersymmetry potential of the LHC and NLC.  For
each machine, it addressed the following questions:

\begin{itemize}
\item Can one identify a signal for new physics?

\item If so, can one tell that the new physics is supersymmetry?

\item If it is supersymmetry, can one distinguish between various
models and measure the underlying parameters?
\end{itemize}

\noindent
The group concluded that the LHC and NLC bring complementary approaches
to the study of the superparticles and their properties.
(In this talk I will not discuss supersymmetry at the Tevatron or LEP.  The
Tevatron and its possible upgrades were also examined by the Supersymmetry
Group at Snowmass \cite{Snowmass:TeV33}.)

\section{THE MSSM}

The Snowmass studies were based on the Minimal Supersymmetric Standard Model
(MSSM), the simplest supersymmetric extension of the ordinary Standard
Model.  The MSSM contains the minimal set of fields:  one superpartner
for every known particle, as well as two Higgs doublets, together with
their associated supersymmetric partners.  The MSSM couplings are assumed
to respect
$R$-parity, which implies that superparticles must be pair produced and
that the lightest supersymmetric particle must be stable.

The MSSM contains two types of parameters.  The first parametrizes the
couplings that are related by supersymmetry.  For example, at tree-level,
supersymmetry requires that the quark-squark-gluino Yukawa coupling be
equal to the quark-quark-gluon and squark-squark-gluon gauge couplings.  
These relations are essential if supersymmetry is to cancel quadratic
divergences.

The supersymmetric parameters include all the gauge and Yukawa couplings
of the Standard Model, plus the Higgsino mass parameter, $\mu$.
The other parameters in the MSSM describe the supersymmetry breaking.
These parameters are soft, in the sense that they do not reintroduce
destabilizing quadratic divergences.  They include Higgs and gaugino
masses, as well as the squark and slepton masses and mixings.  The
supersymmetry-breaking parameters also include a set of trilinear
scalar couplings.  In sum, there are over 100 such parameters,
each of which must be determined by experiment.

Therefore, if supersymmetry is indeed realized by nature, one must

\begin{enumerate}
\item Find the supersymmetric partners;

\item Verify the relations between couplings implied by
supersymmetry; and

\item Measure the soft parameters, to shed light on the mechanism of
supersymmetry breaking.
\end{enumerate}

\begin{figure}[t]
\epsfxsize=3.2in
\hspace*{-0.15in}
\vspace*{-0.3in}
\epsffile{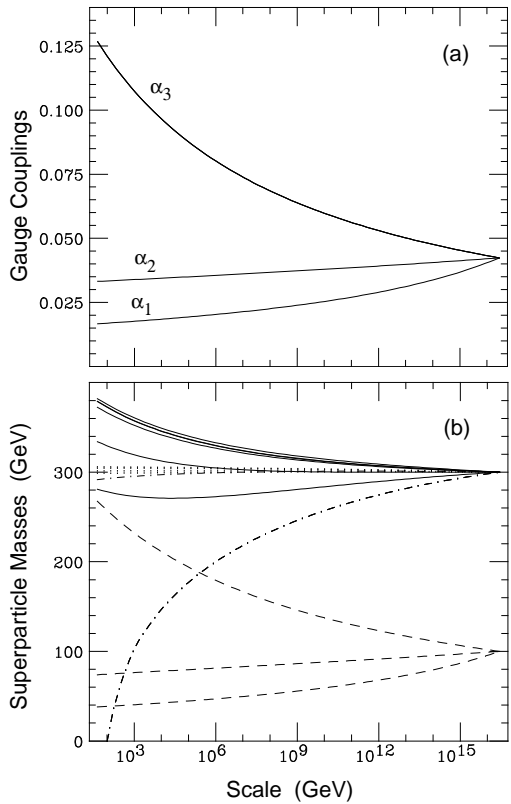}
\caption{(a)  The running gauge couplings unify above $10^{16}$
GeV.  (b) In the mSUGRA scenario, the soft couplings are assumed to unify at
the same scale.  The heavy top quark drives one Higgs mass-squared
negative (dot-dashed line) and triggers electroweak symmetry breaking.}
\label{rng}
\end{figure}

At the Snowmass Workshop, the Supersymmetry Working Group assessed the
ability of the LHC and NLC to carry out this program.  For definiteness,
the group chose to work in the mSUGRA scheme \cite{Snowmass:theory}, in
which the soft supersymmetry-breaking parameters are assumed to unify at
the same scale as the running gauge couplings, $\alpha_1$, $\alpha_2$ and
$\alpha_3$, as shown in Fig.~\ref{rng}.

The mSUGRA scenario involves several bold assumptions, any one of which
might well be wrong \cite{Snowmass:theory}, but it has the tremendous
advantage that it reduces the parameter space to just five new parameters:
a universal scalar mass, $M_0$, a universal gaugino mass, $M_{1/2}$, a
universal trilinear scalar coupling, $A_0$, a Higgs mass, $B\mu$, and the
supersymmetric Higgsino mass, $\mu$.  It is also consistent
with all experimental data to date.

The mSUGRA scenario has the additional attraction that radiative corrections
from the heavy top quark drive electroweak symmetry breaking, as shown in
Fig.~\ref{rng}b.  This allows one to trade $\mu^2$ and $B\mu$ for $M^2_Z$ and
$\tan\beta$, the ratio of Higgs vevs.  Therefore, in the mSUGRA scenario,
the final parameter space is simply:
\begin{displaymath}
M_0,\ \ M_{1/2},\ \ A_0,\ \ \tan\beta,
\end{displaymath}
together with the sign of $\mu$.  The first three parameters are specified
at the unification scale, while $\tan\beta$ is taken at the scale $M_Z$.

In the mSUGRA scenario, the superpartner masses and mixings
are determined
by the above parameters and the supersymmetric renormalization group
evolution to $M_Z$.  The resulting low-energy spectrum tends to
have the following properties:

\begin{itemize}
\item
The weak-scale gaugino mass parameters appear in the ratio
\begin{displaymath}
M_1\ :\ M_2\ :\ M_3\ \sim\ \alpha_1 :\ \alpha_2 :\ \alpha_3;
\end{displaymath}

\item The $\mu$ parameter tends to be large, $|\mu | \gg M_2$.  This
implies that
the lightest neutralino ($\tilde{\chi}^0_1$) is primarily bino ($\tilde B$),
while the second-lightest neutralino ($\tilde{\chi}^0_2$) and the lightest
chargino ($\tilde{\chi}^\pm_1$) are predominantly wino ($\tilde{W}^0,
\tilde{W}^\pm$),
\begin{displaymath}
\tilde{\chi}^0_1\ \sim\ \tilde B,\qquad \quad
\tilde{\chi}^0_2,\ \tilde{\chi}^\pm_1\ \sim\ \tilde W;
\end{displaymath}

\item
Because $|\mu | \gg M_2$, the heavier neutralinos ($\tilde{\chi}
^0_{3,4}$) and chargino ($\tilde{\chi}^\pm_2$) are primarily Higgsino.  
Their masses tend to be somewhat larger than
those of the $\tilde{\chi}^0_{1,2}$ and the $\tilde{\chi}^\pm_1$;

\item
If $M_0 \gg M_{1/2}$, the squarks and sleptons are nearly
degenerate, and are significantly heavier than the $\tilde{\chi}
^0_{1},$ $\tilde{\chi}^0_{2}$ and the $\tilde{\chi}^\pm_1$.
If $M_0 \lsim M_{1/2}$, the squarks are heavier than the
sleptons.
\end{itemize}

\noindent
These relations will be useful when we discuss supersymmetry signatures
at the two machines.

\section{SUPERSYMMETRY REACH}

According to its design specifications, the LHC is a 14 TeV $pp$ collider,
capable of achieving an annual integrated luminosity of 100 fb$^{-1}$. 
For the Snowmass study, it proved sufficient to consider a low-luminosity
option, with just 10 fb$^{-1}$ per year.

Crudely speaking, the LHC provides two intense beams of quarks and gluons which
collide with a center-of-mass energy in the range of a few TeV.  Therefore, a
classic supersymmetry signature is given by
\begin{displaymath}
g\ g\ \rightarrow\ \tg\ +\ \tg,
\end{displaymath}
where one gluino decays to two jets and missing energy and the other decays
to two jets, a lepton, plus missing energy.  This suggests that a supersymmetry
search start with events which contain one lepton, jets and missing energy.

\begin{figure}[t]
\epsfxsize=3.2in
\hspace*{-0.35in}
\vspace*{-.35in}
\epsffile{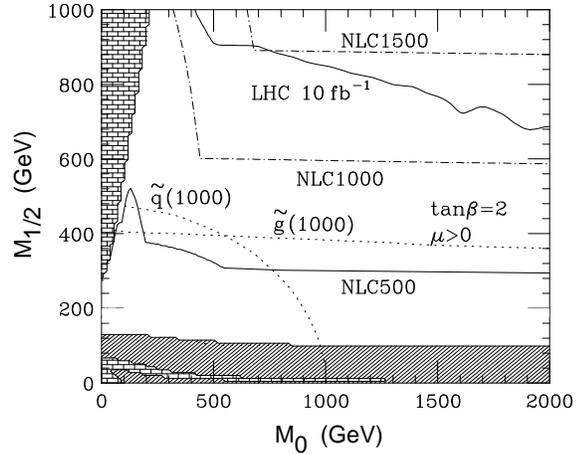}
\caption{The supersymmetry reach of various facilities in the mSUGRA
model, for $\tan\beta=2$, $A_0=0$ and $\mu>0$.  Note that the reach of
a 1.2 $-$ 1.5 TeV NLC is approximately equivalent to that of the LHC.}
\label{reach}
\end{figure}

The results of such a search are shown in Fig.~\ref{reach} \cite{BCPT}. 
The figure shows the $M_0 - M_{1/2}$ plane, for $\tan\beta=2$, $A_0=0$
and $\mu>0$.  Qualitatively speaking, the reach is only weakly sensitive
to this choice.  In the figure, the bricked (hatched) region is excluded
by theoretical (experimental) constraints.

From the figure we see that the LHC gives a gluino reach of between
1.5 and 2.5 TeV, with  just 10 fb$^{-1}$ of luminosity.  In fact, the LHC
subgroup showed that the same cuts and luminosity can be used to detect
squarks and gluinos with masses of up to 1 TeV in supersymmetry {\it without}
$R$-parity \cite{Snowmass:LHC,BCT}.  (With optimized cuts, the LHC can do
even better.)  Taken together, these statements form the basis for the
assertion that the LHC will probe the entire parameter space
of interest for weak-scale supersymmetry. 

The NLC, of course, is at an earlier stage of development, so its specifications
are less certain.  For the Snowmass Workshop, the NLC was {\it defined} to be a
linear $e^+e^-$ collider with a center of mass energy between 500 GeV and 1.5 TeV.  
The luminosity was assumed to be 50 fb$^{-1}$ per year at 500 GeV and up
to 200 fb$^{-1}$ at 1.5 TeV.  It was also assumed that the electrons could
be 80\% polarized.

At the NLC, two classic supersymmetry signatures are chargino and
selectron production,
\begin{displaymath}
e^+e^-\ \rightarrow\ \tilde{\chi}^+_1\ +\ \tilde{\chi}^-_1
\end{displaymath}
and
\begin{displaymath}
e^+e^-\ \rightarrow\ \tilde{e}^+_R\ +\ \tilde{e}^-_R.
\end{displaymath}
In each case, the strategy is to tune the beam energy and search for central
events.  The corresponding reach plot is shown in Fig.~\ref{reach} \cite{BMT}.  
From the
figure we see that the reach of a 1.2 $-$ 1.5 TeV NLC is comparable to that
of the LHC.

It is important to note, though, that at each machine, the quoted reach is
achieved through different channels.  The LHC reach relies on gluinos,
while that of the NLC derives from charginos and sleptons. In the mSUGRA scenario,
the lighter mass of the charginos and sleptons tends to compensate the lower
energy of the NLC machine.  While this is a reasonable guess, it is by no means
assured that supersymmetry works out this way.

\section{SUPERSYMMETRY ANALYSIS}

The issue of reach is important, but it is just the first step in the experimental
analysis of supersymmetry.  One would like to go further and test the expected
relations
between the supersymmetric parameters and measure the soft supersymmetry-breaking
parameters.  In this way one can establish whether an observed signal is actually
supersymmetry and examine whether it is compatible with mSUGRA or some
other supersymmetric framework.  To see what can be done, one needs to work with
a complete model in which one can analyze a variety of experimental signatures.

At Snowmass the Supersymmetry Working Group studied representative points in the
mSUGRA parameter space, five for each machine.  One point, the Snowmass comparison
point, was chosen to be the same for each:
$M_0 = 200$ GeV, $M_{1/2} = 100$ GeV, $A_0 = 0$,
$\tan\beta = 2$ and sign($\mu) = -1$.  For this choice, $m_{\tilde{g}} = 298$ GeV,
with the first two generations of squarks about 20~GeV heavier.  The slepton masses
range between 206 and 216 GeV.  The $\tilde{\chi}^\pm_1$ and $\tilde{\chi}^0_2$
masses are 96 and 97 GeV, while the $\tilde{\chi}^0_1$ mass is 45 GeV.  The
heavier charginos and neutralino masses are between 260 and 270 GeV.  The lighter
stop (sbottom) mass is 264 (278) GeV, while the corresponding heavier states are
more massive than the gluino.

Note that at the comparison point, all the superparticle masses are rather light.
This affects the analysis in at least two ways.  The light masses help the LHC because
they increase the supersymmetry cross sections.  This allowed the LHC subgroup
to make hard cuts to isolate pure samples of supersymmetric events.  The light
masses help the NLC as well because they decrease the energy
requirements.  In particular, at the comparison point, a 500 GeV NLC is
able to access much of the supersymmetric physics.  Note too that in this scenario,
the Higgs mass is just 68 GeV, so the Higgs would probably have been discovered
after Snowmass but before SUSY97!

The analyses at the comparison point take advantage of the following decay chains.

\begin{enumerate}
\item
The lightest neutralino is primarily bino and the lightest chargino is
mostly wino, so the decay into a charged lepton $(\ell = e, \mu)$,
\begin{displaymath}
\tilde{\chi}^\pm_1 \ \rightarrow\ \ell^\pm\ +\ \et,
\end{displaymath}
has a branching ratio of 22\%.  Likewise, the decay into jets,
\begin{displaymath}
\tilde{\chi}^\pm_1 \ \rightarrow\ 2\ {\rm jets}\ +\ \et,
\end{displaymath}
has a branching ratio of 66\%.

\item
The second-lightest neutralino is primarily wino.  Its decay into
charged leptons
\begin{displaymath}
\tilde{\chi}^0_2 \ \rightarrow\ \ell^+\ +\ \ell^-\ +\ \et
\end{displaymath}
has a branching ratio of 33\%.

\item
Because $m_{\tilde g} > m_{\tilde b}$, but $m_{\tilde g} < m_{\tilde q}$
for the first two generations, the decay chain
\begin{eqnarray*}
\tilde{g} &\rightarrow &\tilde{b}\ +\ b \\
&& \ \decay{5}\ \ \tilde{\chi}^0_2\ +\ b \\
&& \quad \quad \, \decay{5}\ \ \tilde{\chi}^0_1\ +\ \ell^+\ +\ \ell^-
\end{eqnarray*}
has a branching ratio of 25\%.  This gives rise to spectacular signatures
with multiple $b$ jets, leptons and missing energy.
\end{enumerate}

\subsection{LHC Strategy}
\vspace{0.1in}

At the LHC, the analysis of supersymmetry is likely to be quite complicated
because all the superpartners are produced at once.  Indeed, it is often
said that at the LHC, the ``background to supersymmetry is supersymmetry."
This suggests a systematic approach, perhaps along the following lines:

\begin{enumerate}
\item Determine the supersymmetry scale;

\item Study the clean decay chains; and

\item Test specific models by performing global fits to the data.
\end{enumerate}

\noindent
At Snowmass, the LHC subgroup carried out such an analysis for the comparison
point \cite{Snowmass:LHC,PRD:LHC}.

\begin{figure}[tb]
\epsfxsize=3.1in
\hspace*{-0.1in}
\vspace*{-.35in}
\epsffile{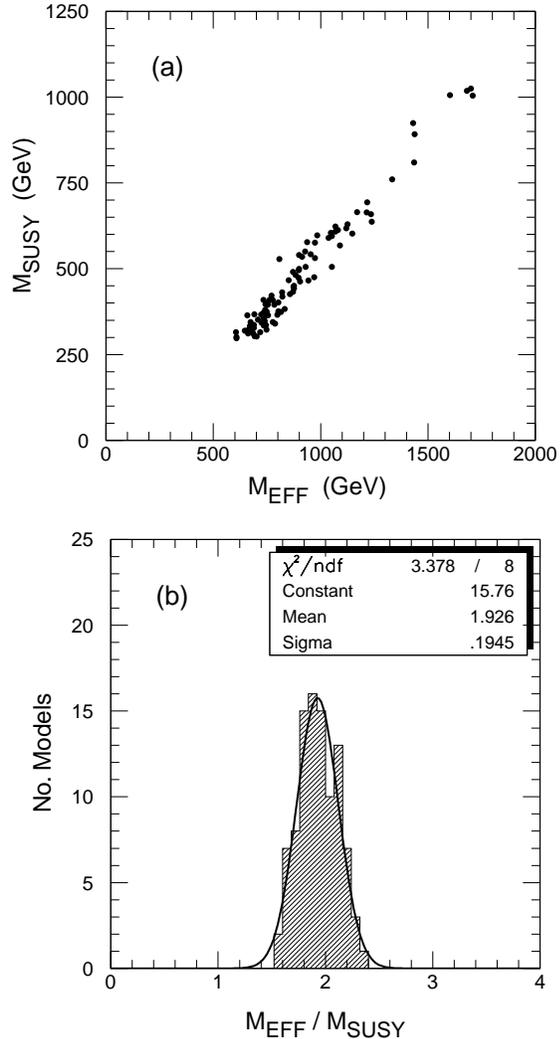}
\caption{(a)  $M_{\rm SUSY}$ versus the peak of the $M_{\rm EFF}$
distribution, for 100 mSUGRA models.  (b)  The ratio $M_{\rm SUSY}/
M_{\rm EFF}$ is constant to within 10\%.}
\label{meff}
\end{figure}

The first step is to determine the overall scale of supersymmetry, which can
be defined as the smaller of the squark or gluino mass,
\begin{equation}
M_{{\rm SUSY}}\ =\ {\rm min}(m_\tg, m_{\tilde u}).
\end{equation}
The LHC subgroup proposed the following technique.  One first selects a sample
of events with at least four jets, no leptons and substantial missing energy.
One then computes the scalar sum of the transverse momenta of the four hardest
jets, plus the missing energy,
\begin{equation}
M_{\rm EFF}\ =\ \sum_{i=1}^{4}\ |p_{{\rm T\, jet,}\, i}|\ +\ \et,
\end{equation}
for each event.  It turns out that the peak in the $M_{\rm EFF}$ distribution
is closely correlated with $M_{{\rm SUSY}}$, at least in the mSUGRA scenario
\cite{Snowmass:LHC,PRD:LHC,Paige}.

The LHC subgroup illustrated this procedure for 100 mSUGRA models (with the same
Higgs mass).  The results are shown in Fig.~\ref{meff}, where one sees that the
ratio $M_{\rm EFF}/M_{{\rm SUSY}}$ is essentially constant.  Indeed, one finds
$M_{\rm EFF}/M_{{\rm SUSY}} = 1.9$ to within 10\%, at least within the class of
mSUGRA models discussed here \cite{Snowmass:LHC,PRD:LHC,Paige}.

The LHC subgroup then turned its attention to
the clean decay chains.  Such an analysis is easiest if one starts with a
relatively pure sample of supersymmetry events.  For the Snowmass comparison point,
there are 13.5 {\it million} supersymmetry events per 10 fb$^{-1}$ of luminosity.  
This permits hard cuts to isolate pure samples of supersymmetric decays.

As discussed above, at the Snowmass point, the decay $\tilde{g} \rightarrow
b \,+\, \bar b\, +\, \ell^+ +\, \ell^- +\, \et$ has a branching ratio of 25\%.  
With 10 fb$^{-1}$ of luminosity and a tagging efficiency of 60\% (and $c$
misidentification of 10\%) per $b$-jet, this gives rise to 272k events which
contain four $b$ jets and two pairs of opposite-sign same-flavor leptons.
Moreover, if one of the $\tilde{\chi}^0_2$ is allowed to decay hadronically,
there are another 694k events with four $b$ jets, two non-$b$ jets and one
pair of opposite-sign same-flavor leptons.  This suggests that the supersymmetry
sample be selected to contain four or more jets, with at least two
tagged as $b$-jets and one or more pairs of opposite-sign same-flavor leptons.

\begin{figure}[tb]
\epsfxsize=3.1in
\hspace*{-0.1in}
\vspace*{-.45in}
\epsffile{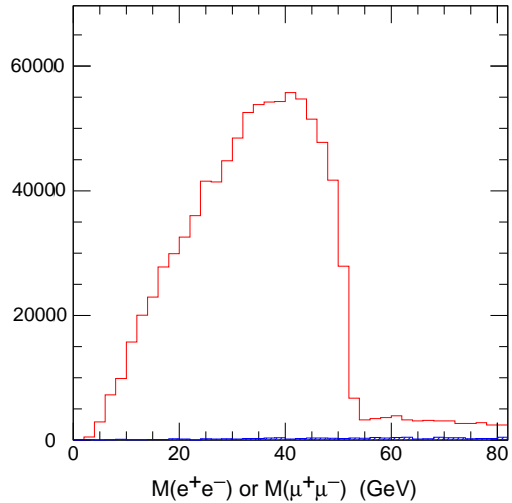}
\caption{The dilepton mass spectrum from LHC events with four or more jets,
with at least two tagged as $b$-jets, one or more pairs of opposite-sign
same-flavor leptons, and 10 fb$^{-1}$ of integrated luminosity.}
\label{chi2-chi1}
\end{figure}

The LHC subgroup collected such a sample and plotted the dilepton mass
distribution, as shown in Fig.~\ref{chi2-chi1}
\cite{Snowmass:LHC,PRD:LHC,Soderqvist}.  
By fitting the sharp edge of the distribution, they determined
that the mass difference $m_{ \tilde{\chi}^0_2} - m_{\tilde{\chi}^0_1}$ can
be measured to $\pm 50$ MeV!  
This incredible precision follows from the fact that the measurement is
systematics limited and the dilepton mass can be calibrated against
$M_Z$.

The subgroup then used the technique of partial reconstruction, first
developed for $D^* \rightarrow D^0 \pi$ decays \cite{CLEO}, to show
that the difference
$m_{\tilde{b}} - m_{\tilde{\chi}^0_2}$ can be measured to $\pm 2$
GeV \cite{Snowmass:LHC,PRD:LHC,Yao}.  The technique goes as follows.   One
first selects events near the edge of the $\ell^+\ell^-$ mass distribution.
Since the leptons come from the decay chain
\begin{eqnarray*}
\tilde{g} &\rightarrow &\tilde{b}\ +\ b \\
&& \ \decay{5}\ \ \tilde{\chi}^0_2\ +\ b \\
&& \quad \quad \, \decay{5}\ \ \tilde{\chi}^0_1\ +\ \ell^+\ +\ \ell^-,
\end{eqnarray*}
the events near the edge originate in decays in which the $\tilde{\chi}^0_1$
and the $\ell^+\ell^-$ pair are at rest in the rest frame of the $\tilde{\chi}
^0_2$.  If one then {\it assumes} some mass for the $\tilde{\chi}^0_1$,
one can boost back to the lab and reconstruct the entire event.  About 6000
gluino and sbottom events can be reconstructed in this way per 10 fb$^{-1}$
of luminosity at the LHC.

\begin{figure}[tb]
\epsfxsize=3.1in
\hspace*{-0.1in}
\vspace*{-.45in}
\epsffile{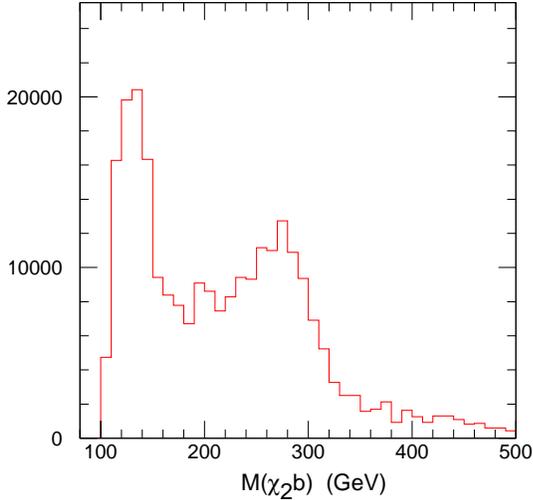}
\caption{The partially-reconstructed bottom squark mass at the LHC with
10 fb$^{-1}$ of integrated luminosity.}
\label{bsquark}
\end{figure}

\begin{figure}[tb]
\epsfxsize=3.1in
\hspace*{-0.1in}
\vspace{-.42in}
\epsffile{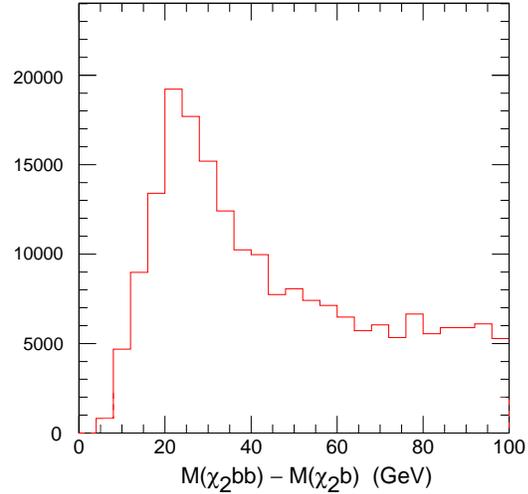}
\caption{The difference between the gluino and the bottom squark mass
using the partial reconstruction technique at the LHC with 10 fb$^{-1}$
of integrated luminosity.}
\label{gluino-bsquark}
\end{figure}

Using this technique and assuming a value for $m_{\tilde{\chi}^0_1}$, one can
measure the mass of the $\tilde{b}$, as well as the mass difference $m_{\tg} -
m_{\tilde{b}}$, as shown in Figs.~\ref{bsquark} and \ref{gluino-bsquark}.  By 
varying the value of $m_{\tilde{\chi}^0_1}$, one can show that $m_{\tg} 
- m_{\tilde{b}}$ is essentially independent of the assumption, up to an error
of $\pm 2$ GeV.  One also finds that
\begin{eqnarray*}
\Delta m_{\tilde{b}} &=& 1.5\,[\,m_{\tilde{\chi}^0_1}
({\rm assumed}) - m_{\tilde{\chi}^0_1}({\rm true}) \,] \\
&& \pm\ 3\ {\rm GeV.}
\end{eqnarray*}
Therefore once $m_{\tilde{\chi}^0_1}$ is known, the $\tilde{\chi}^0_2$
and $\tilde b$ masses are determined to the percent level!

The final step in the analysis is to try to determine the soft
supersymmetry-breaking parameters.  At the LHC, this might be accomplished
by performing a global fit to the data.  If one is lucky, and the soft
parameters originate in some simple model, such as mSUGRA or gauge mediation,
the fit will agree with the data.  If not, and all 100 parameters need to be
measured independently, the global fit will not work and one will be forced
to try something else.

For the case at hand, the global fit works very well.  The measured inputs
\begin{eqnarray*}
m_{\tilde{\chi}^0_2} - m_{\tilde{\chi}^0_1}
&=& 52.36 \pm 0.05\ {\rm GeV} \\
m_{\tg} - m_{\tilde{b}}
&=& 20.3 \pm 2.0 \ {\rm GeV} \\
m_h
&=& 68.3 \pm 3 \ {\rm GeV}
\end{eqnarray*}
(where the Higgs mass measurement comes from LEP, and the error is purely
theoretical), imply
\begin{eqnarray*}
M_0
&=&
200^{+13}_{-8}\ {\rm GeV} \\
M_{1/2}
&=&
99.9 \pm 0.7\ {\rm GeV} \\
\tan\beta
&=&
1.95 \pm 0.05 \\
{\rm sign}(\mu)
&=& {\rm determined} \\
A_0
&>& -400 \ {\rm GeV}.
\end{eqnarray*}

The fit works as follows:  Since the $\tilde{\chi}^0_2$ and $\tilde{\chi}^0_1$
are effectively wino and bino, the value of $M_{1/2}$ is fixed by
$m_{\tilde{\chi}^0_2} - m_{\tilde{\chi}^0_1}$.  This, in turn, gives
$m_{\tilde{\chi}^0_1}$ in the mSUGRA scenario.  Partial reconstruction
then determines $m_{\tilde {b}}$, hence $M_0$.  Finally,
the value of $m_h$ is sufficient to fix $\tan\beta$.

Given this global fit, one can carry out many cross checks to test the consistency
of the solution.  For example, once $M_{1/2}$ is known, the gluino mass can be
predicted and compared with experiment.  Similarly, the event rates and branching
ratios can be used to test the fit.

\subsection{NLC Strategy}
\vspace{0.1in}

\begin{figure}[tb]
\epsfxsize=3.1in
\hspace*{-0.1in}
\vspace{-.37in}
\epsffile{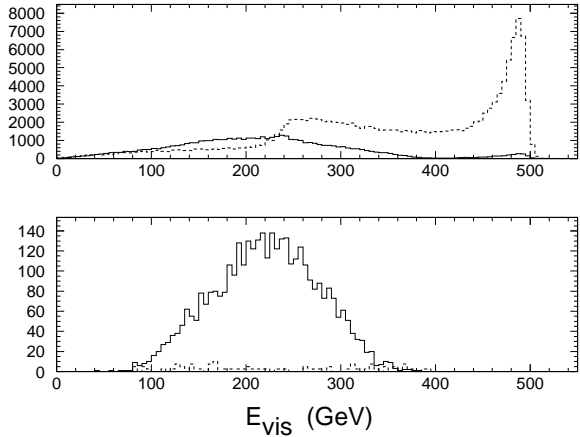}
\caption{The distribution of total visible energy for supersymmetry (solid line)
and $WW$ (dashed line) events, before and after cuts.  The
signal to background ratio is 30:1 after the cuts.}
\label{sample}
\end{figure}

Supersymmetry studies at the NLC exploit the facts that the beam energy is 
adjustable, the electron polarization can be varied, and signal events are
central.  The adjustable energy helps one to produce just the particles of
interest, while the polarization and event shape help to separate the signal
from the background.  Operation near threshold permits accurate mass
measurements, while precise knowledge of the collision energy greatly
facilitates the reconstruction of events with undetected particles in the final
state.  These facts are illustrated in Fig.~\ref{sample}, where a series of
cuts at $\sqrt s = 500$ GeV and 80\% polarization have been used to produce a
supersymmetric sample with 30:1 ratio of signal to background.

The JLC \cite{JLC} group has pioneered a program in which they have shown that
chargino, neutralino and slepton masses can be accurately measured at a linear
$e^+e^-$ collider.  The JLC studies ignored cascade decays and assumed an
electron beam polarization of 95\%.  At Snowmass the NLC group refined this
work by including the effects of cascade decays and taking the electron
polarization to be a more conservative 80\% \cite{Snowmass:NLC}.  
(The precision measurements
that might be possible at the NLC with cascade decays have also been
examined in \cite{BMT}.)

\begin{figure}[tb]
\epsfxsize=3.1in
\hspace*{-0.1in}
\vspace{-.37in}
\epsffile{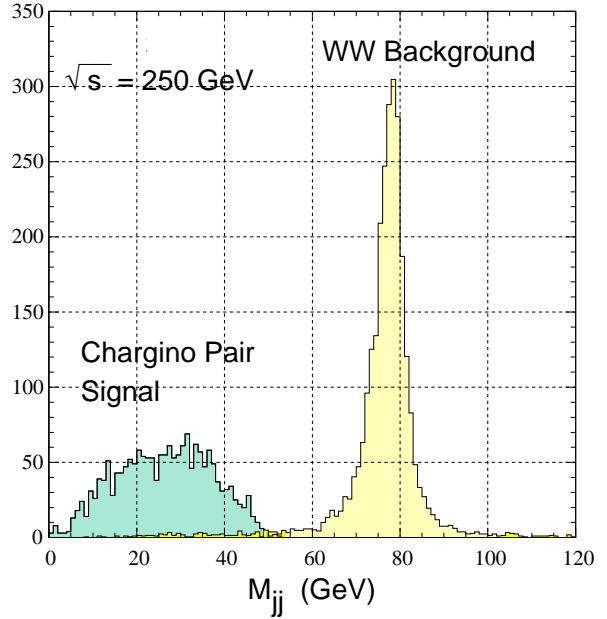}
\caption{The dijet distributions from chargino pair production and the
Standard-Model $WW$ background at the NLC.}
\label{chi-dijet}
\end{figure}

The NLC subgroup began its study of the supersymmetry comparison point by
running its machine at $\sqrt s = 250$ GeV.  At this energy just the lightest
chargino and neutralino are produced.  Chargino pair production,
\begin{displaymath}
e^+e^-\ \rightarrow\ \tilde{\chi}^+_1\ +\ \tilde{\chi}^-_1,
\end{displaymath}
where one chargino decays leptonically, and the other hadronically,
$\tilde{\chi}^\pm_1 \rightarrow$ 2 jets + $\tilde{\chi}^0_1$,
gives rise to a striking signal in the jet-jet invariant mass distribution.
Indeed, as shown in Fig.~\ref{chi-dijet}, this signal can be readily separated
from the $WW$ background.

\begin{figure}[tb]
\epsfxsize=3.1in
\hspace*{-0.1in}
\vspace{-.55in}
\epsffile{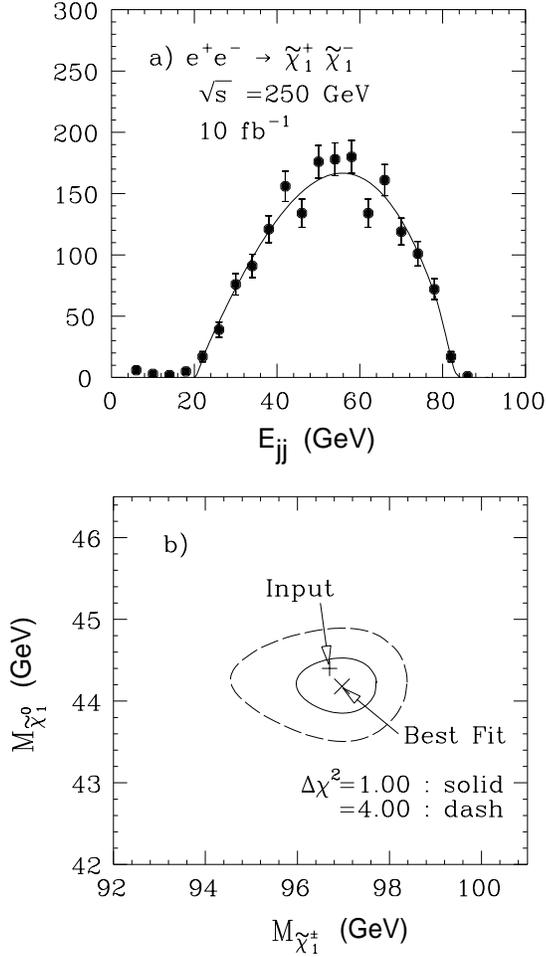}
\caption{The dijet energy distribution, and the masses obtained by a fit
to its endpoints.}
\label{chimass}
\end{figure}

A fit to the endpoints of the dijet mass distribution gives a measurement of
$m_{\tilde {\chi}^0_1}$ and $m_{\tilde{\chi}^\pm_1}$, as shown in Fig.~\ref
{chimass}.  With 20 fb$^{-1}$ of data, the NLC subgroup recovered the input
values, with an error
of 1\% on both measurements \cite{Snowmass:NLC}.

\begin{figure}[tb]
\epsfxsize=3.0in
\epsffile{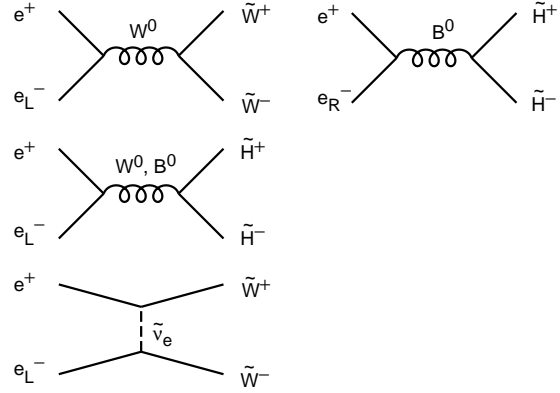}
\caption{Left-polarized electrons can produce the $\tilde {\chi}^\pm_1$ whether
it is wino or Higgsino, whereas right-polarized electrons can only produce it
if it is Higgsino.  ($B$-$W$ mixing is small at NLC energies.)}
\label{pol}
\end{figure}

The polarization of the beam can be used to determine the wino/Higgsino
content of the $\tilde{\chi}^\pm_1$.  The key point is that left-polarized
electrons can produce the $\tilde{\chi}^\pm_1$ whether it is wino or Higgsino,
whereas right-polarized electrons can only produce it if it is Higgsino (see
Fig.~\ref{pol}). Therefore, if the left-polarized cross section, $\sigma_L$, 
is much greater than the right-polarized cross section, $\sigma_R$, the
$\tilde{\chi}^\pm_1$ is primarily wino.  This analysis is independent of
any assumptions about the nature of the soft supersymmetry-breaking terms.

For the case at hand, the NLC subgroup showed that with 20 fb$^{-1}$ of data,
the polarized cross sections can be measured to an accuracy of less than 2\%.
From the cross sections they then inferred that the lightest chargino is indeed
a wino, and that the mass of the $t$-channel sneutrino must be less than 250
GeV \cite{Snowmass:NLC}.

\begin{figure}[tb]
\epsfxsize=3.1in
\hspace*{-0.1in}
\vspace{-.35in}
\epsffile{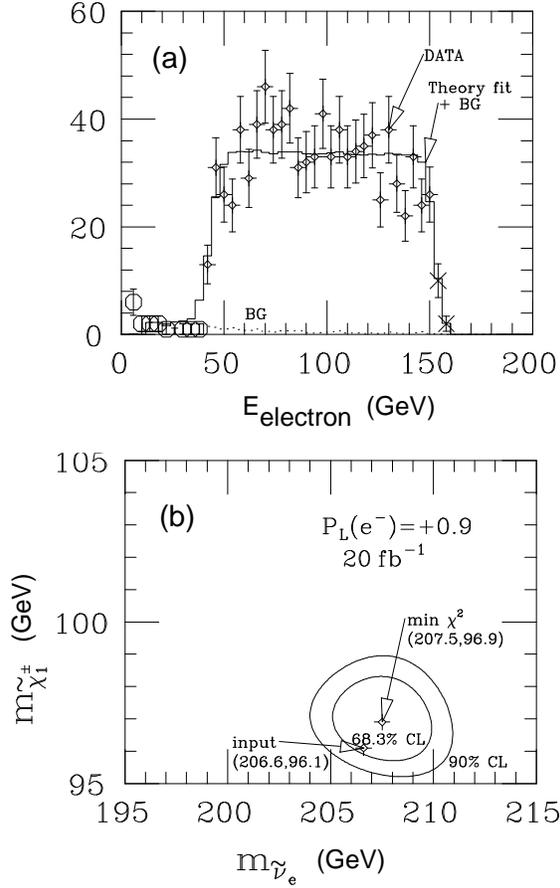}
\caption{(a) The electron energy distribution in the process $e^+e^-
\rightarrow \tilde{\nu}_e + \tilde{\nu}_e \rightarrow e^+ + e^- +
\mu^\pm + 2\ {\rm jets} + \et$.
(b) The fit that determines the chargino and sneutrino masses.}
\label{sneut}
\end{figure}

Based on this information, the subgroup proposed increasing the NLC energy
to 500 GeV, which is sufficient to pair-produce all sleptons.  Indeed,
sneutrino pair production,
\begin{displaymath}
e^+e^- \ \rightarrow\ \tilde{\nu}_e\ +\ \tilde{\nu}_e^*,
\end{displaymath}
followed by the decays
\begin{eqnarray*}
\tilde{\nu}_e\  &\rightarrow & \tilde{\chi}^+_1\ +\ e^-  \\
&& \ \decay{5}\ \ \mu^+\ +\ \et \\
\tilde{\nu}_e^* &\rightarrow & \tilde{\chi}^-_1\ +\ e^+  \\
&& \ \decay{5}\ \ {\rm 2\ jets}\ +\ \et
\end{eqnarray*}
(together with their charge conjugates), gives rise to the electron
energy distribution shown in Fig.~\ref{sneut}.  The endpoints of the
distribution determine $m_{\tilde{\nu}_e}$ and $m_{\tilde{\chi}
^\pm_1}$.  With 25 fb$^{-1}$ of data and 80\% left-handed polarization,
the NLC ZDR study \cite{ZDR} quotes the following masses,
\begin{eqnarray*}
m_{\tilde{\nu}_e} &=& 207.5 \pm 2.5 \ {\rm GeV} \\
m_{\tilde{\chi}^\pm_1} &=& 97.0 \pm 1.2 \ {\rm GeV,}
\end{eqnarray*}
compared with the input values of 206.6 and 96.1 GeV, respectively.

At Snowmass the NLC subgroup studied right-handed selectron production,
\begin{displaymath}
e^+e^- \ \rightarrow\ \tilde{e}^+_R\ +\ \tilde{e}^-_R,
\end{displaymath}
for $\sqrt s = 500$ GeV.  They selected events where each selectron
decays into an electron and a $\tilde{\chi}^0_1$.  For such events, the
endpoints of the electron distribution determine the masses of the $\tilde{
e}^\pm_R$ and the $\tilde{\chi}^0_1$.  They found that the fit values of
the masses did not coincide exactly with the inputs because of backgrounds
from other decays.  Nevertheless, they stated that this effect is correctable,
and estimated an error of $\pm 1\%$ on $m_{\tilde{e}^\pm_R}$, with 20 fb$^{-1}$
of luminosity and 80\% right-handed polarization \cite{Snowmass:NLC}.  
(A similar technique
works for the $\tilde{\mu}^\pm_R$.  Measurement of the $\tilde{e} ^\pm_L$
mass is difficult because left-polarized electrons have
significant Standard-Model backgrounds.  The NLC subgroup quoted an
uncertainty of $\pm 7\%$ using the six-electron final state.)

The four measurements $m_{\tilde{\chi}^\pm_1}$, $m_{\tilde{\chi}^0_1}$,
$\sigma_R(\tilde{e}^+_R \tilde{e}^-_R)$ and $\sigma_R(\tilde{\chi}^+_1
\tilde{\chi}^-_1)$ are sufficient to make a {\it model-independent}
measurement of $\mu$, $M_1$, $M_2$ and $\tan\beta$, without any assumptions
about the mechanism of supersymmetry breaking \cite{JLC}.  This allows
a direct test of the gaugino mass relation between $M_1$ and $M_2$.
Alternatively, one can assume this relation and then determine $M_2$,
$\mu$ and $\tan\beta$.  The NLC subgroup took this second approach and
determined the parameter $M_2$ to $\pm 1.5\%$
\cite{Snowmass:NLC}.

These results, together with the measurements of the slepton masses, led the
NLC subgroup to predict an average squark mass of $322 \pm 7$ GeV in the
mSUGRA scenario.  They therefore suggested increasing the NLC energy to 800
and 1000 GeV.  At these energies they would be able to measure the squark
masses, as well as those of the heavier chargino, neutralinos and Higgs
bosons \cite{Snowmass:NLC}.

\begin{figure}[tb]
\epsfxsize=3.1in
\hspace*{-0.2in}
\vspace{-.35in}
\epsffile{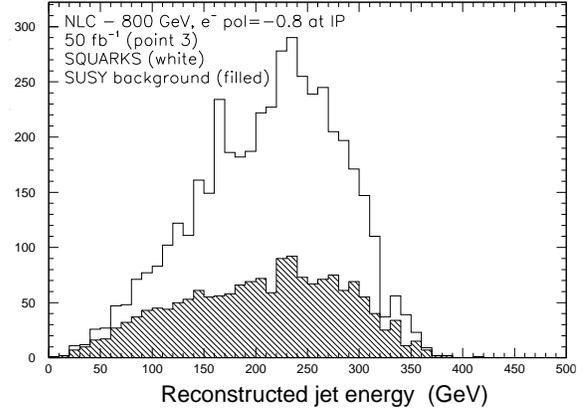}
\caption{The jet energy distribution from squarks (white) and other supersymmetric
background (filled) for events with two $b$ jets at the NLC.}
\label{squarks}
\end{figure}

The NLC subgroup measured the top and bottom squark masses using an event sample
at $\sqrt s = 800$ GeV.  Since these squarks decay predominantly through
\begin{eqnarray*}
\tilde{t} \rightarrow b\ +\ \tilde{\chi}^+_1 \\
\tilde{b} \rightarrow b\ +\ \tilde{\chi}^0_1,
\end{eqnarray*}
they required each event to contain two $b$ jets.  The resulting jet energy
spectrum provides an indication of the third-generation squark mass.  
The jet energy distribution is shown in Fig.~\ref{squarks}, at $\sqrt s = 800$ GeV,
for 50 fb$^{-1}$ of luminosity with 80\% right-handed polarization.  By
fitting the endpoints, the
subgroup measured an average third-generation squark mass of 307 GeV, plus
or minus 10\% \cite{Snowmass:NLC}.

The NLC has the great advantage that the supersymmetric masses and cross sections
can be measured in a systematic, model-independent fashion.  Once one has determined
the masses, one can test whether they fit any particular hypothesis about the origin
of supersymmetry breaking.  For example, measurement of the selectron and smuon
masses directly tests the flavor dependence of the slepton masses.  

At Snowmass
the NLC subgroup fit their measurements to the mSUGRA hypothesis.  They found that
the best fit corresponded to the input parameters, with the following errors
\cite{Snowmass:NLC}:
\begin{eqnarray*}
\delta M_0 &=& {}^{+2.7}_{-2.7}\ {\rm GeV}\\
\delta M_{1/2} &=& {}^{+2.5}_{-1.0}\ {\rm GeV}\\
\delta \tan\beta &=& {}^{+0.17}_{-0.31}\\
{\rm sign}(\mu) &=& {\rm determined,}
\end{eqnarray*}
with no constraint quoted on $A_0$.  (Their fit to $\tan\beta$ would have been
better if they had used the anticipated result from LEP.  Note that neither
the NLC nor the LHC were able to measure $A_0$ because it does not affect
the weak-scale phenomenology.)  As with the LHC, the
predictions from this fit can be cross tested through many other
measurements.  For example, finding squarks at the expected mass would provide
striking support of the mSUGRA scenario.

\begin{figure}[tb]
\epsfxsize=3.1in
\hspace*{-0.1in}
\vspace{-.4in}
\epsffile{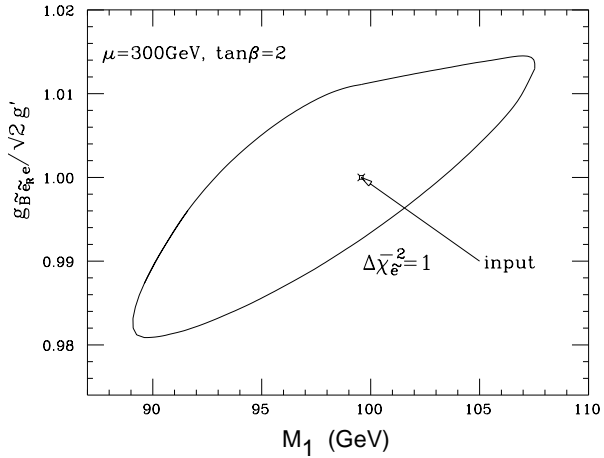}
\caption{The ratio of the bino-selectron-electron Yukawa coupling to the
hypercharge gauge coupling, with 100 fb$^{-1}$ of luminosity at the
500 GeV NLC (assuming 100\% polarization).}
\label{susytest}
\end{figure}

The NLC also offers the exciting possibility of testing the relationships implied
by supersymmetry itself.  An example of such a test is provided by the reaction
\cite{Nojiri}
\begin{displaymath}
e^+e^-\ \rightarrow\ \tilde{e}^+_R\ +\ \tilde{e}^-_R.
\end{displaymath}
The production cross section involves an $s$-channel $B$ and a $t$-channel
$\tilde B$.  At tree level, supersymmetry relates the bino-selectron-electron
Yukawa coupling $g_{\tilde B \tilde{e}_R e}$ to the hypercharge gauge coupling,
$g'$.  With 100 fb$^{-1}$ of luminosity, it may be possible to measure the
ratio $g_{\tilde{B} \tilde{e}_R e} /\sqrt{2} g'$ to better than 2\%, as shown 
in Fig.~\ref{susytest} (for a particular set of parameters).
This test is so precise that it begins to be sensitive to radiative
corrections, which, in turn, are sensitive to the masses of superparticles
which may not be kinematically accessible.  Indeed, it might even be
possible to extract an estimate of the squark mass scale from
this ratio \cite{susycorrections}.

\section{CONCLUSIONS}

In this talk I have tried to give an indication of the types of supersymmetry
measurements that will be possible at the LHC and NLC.  If the accelerators
and detectors work as designed -- and supersymmetry is discovered -- I have
no doubt that the actual measurements will be better than anything I have
reported here.  After all, nothing beats real experimentalists with
real data.

At Snowmass, it is fair to say that the Supersymmetry Working Group reached
the following consensus.

\begin{itemize}
\item
The LHC is an excellent machine for supersymmetry because of its enormous
discovery reach.  It can perform some precision measurements and can confirm
or exclude specific hypotheses on the origin of supersymmetry breaking.

\item The NLC is an excellent machine for supersymmetry because it
permits a systematic, model-independent determination of the
supersymmetry parameters.  It has a discovery reach that is limited by
the available center-of-mass energy.  However, it can perform
many precision measurements for the superparticles within its reach.  
\end{itemize}

I think that experiments at the LHC and NLC do indeed provide complementary
information.  They will discover weak-scale supersymmetry, if it exists,
and will carry out detailed measurements of the superparticle properties.   
The information they provide will help unravel the mysteries of supersymmetry
and supersymmetry breaking and will further our understanding of how
electroweak gauge symmetry is broken.

I would like to thank the co-conveners of the Supersymmetry Working Group,
Uriel Nauenberg, Andy White, and especially, Xerxes Tata, for their close
collaboration during the Snowmass Workshop.
I would also like to thank the subgroup organizers
and participants for the skill and enthusiasm with which they carried out
the work on which this report is based.

\end{document}